# Electrocaloric effect in $Ba_{0.2}Ca_{0.8}Ti_{0.95}Ge_{0.05}O_3$ determined by a new pyroelectric method

B. Asbani[1], J-L. Dellis[1], Y. Gagou[1], H. Kaddoussi[1], A. Lahmar[1], M. Amjoud[2], D. Mezzane[2], Z. Kutnjak[3], M. El Marssi[1,(a)]

[1] Laboratoire de Physique de la Matière Condensée, Université de Picardie, 33 rue Saint-Leu, 80039 Amiens Cedex, France
[2] LMCN, F.S.T.G. Université Cadi Ayyad, BP 549, Marrakech, Morocco.
[3] Jozef Stefan Institute, Jamova cesta 39, 1000 Ljubljana, Slovenia

**Abstract** – The present letter explores the electrocaloric effect (ECE) in the lead free oxide $Ba_{0.8}Ca_{0.2}Ti_{0.95}Ge_{0.05}O_3$ ceramics (BCTG). The electrocaloric responsivity $\xi = (\frac{dT}{dE})$ was determined by two different methods using the Maxwell relationship $\xi \propto \left(\frac{\partial P}{\partial T}\right)_E$. In a first well-known indirect method, P-E hysteresis loops were measured in a wide temperature range from which the pyroelectric coefficient $p_E = \left(\frac{\partial P}{\partial T}\right)_E$ and thus $\xi$ were determined by derivation of P(T,E) data. In the second novel method the pyroelectric coefficient $p_E$ and consequently the electrocaloric responsivity $\xi$ was determined by direct measurements of the pyroelectric currents under different applied electric fields. Within the experimental error good agreement was obtained between two methods with $\xi = 0.18 \pm 0.05$ $10^{-6}$ K.m.V$^{-1}$ was obtained at about 410 K.

The electrocaloric effect is a coupling of electrical and thermal properties that results in an adiabatic temperature change $\Delta T$ in response to an externally applied electric field $\Delta E$ [1]. Nowadays, refrigeration based on the ECE is a new solid-state cooling technology that is rapidly developing. The electrocaloric effect has a significant technological importance, enabling the production of high-performance solid-state cooling devices for a broad range of applications, such as temperature regulation for sensors, electronic devices, and on-chip cooling devices [2]. It presents the advantage that the application of an electric field requires less complex components and control [3–5] than other technologies. Ideally, good electrocaloric materials should show high temperature and entropy changes as a function of electric field and temperature. In the ferroelectric materials, the application of an electric field induces a large polarization change which is particularly suitable for this purpose. In 2006, Mischenko *et al.* [3] first demonstrated a 'giant' electrocaloric effect in PZT thin films at the phase transition temperature triggering a renewed interest in the electrocaloric effect. Subsequently, the ECE has been reported for many different ferroelectric materials such as thick and thin films [6–9], polymers [4,10,11], ceramics [12–15], and single crystal [16–19]. There were a few studies on ECE made with direct measurement on $\Delta T$, due to the difficulties to undertake these measurements especially for films [20,21]. Most experimental studies were based on the Maxwell relationships in which the polarization versus temperature $T$ under different electric fields is measured. Usually the standard method consists to make P(E) hystersis loops at different temperatures. Then the pyroelectric coefficient $p_E$ is computed from the upper branches of the hysteresis loops [3] allowing the $\Delta T$ determination through the integration:

$$\Delta T = -\int_{E_1}^{E_2} \left(\frac{T}{\rho . c_p}\right) \left(\frac{\partial P}{\partial T}\right)_E dE \quad (1)$$

where $E_1$ and $E_2$ define the branche boundaries, $\rho$ is the density, $c_p$ is the specific heat capacity and the integrand is the electrocaloric responsivity:

$$\xi = \left(\frac{\partial T}{\partial E}\right)_T \quad (2)$$

In the current effort to find lead-free materials for the electronics, Moya *et al.*[22] recently reported a giant responsivity $\xi$ of about 2.2 $10^{-6}$ K.m.V$^{-1}$ in $BaTiO_3$ single crystal near the sharp first-order phase transition with a large latent heat. Our study presents the results obtained on a lead-free ceramic, based on $BaTiO_3$ and doped with calcium and germanium. Moreover, we compared the standard method with a new one in which $p_E$ is measured directly.

Dense BCTG polycrystalline ceramics were elaborated using high purity raw materials of $BaCO_3(99\%)$, $CaCO_3(98.5\%)$, and $TiO_2(99.8\%)$, $GeO_2(99.99\%)$. The details of preparation method are described in Ref.[23]. The measurements were performed on pellet of diameter about 5.8 mm and thickness of about 0.82 mm. In order to measure the pyroelectric coefficient, conductive electrodes (Au) were deposited on the samples by sputtering, using a SCD050 device and had a thickness of about 50 nm. The voltages applied to polarize the sample and the current across it were

[(a)]E-mail: mimoun.elmarssi@u-picardie.fr

generated and measured, respectively by a Keithley 2635B SourceMeter. The ceramic temperature was controlled using a Linkam TMS 600.

Figure 1(a) shows the X-ray diffraction pattern recorded at room temperature for the sintered BCTG sample. It is evident from the XRD patterns that the sample exhibits a pure polycrystalline perovskite structure without any impurity phase within the detection limits of our instrument. Sharp and well defined diffraction peaks indicate that ceramics have long range crystalline order. This also clearly indicates that Ge is dissolved in the BCT lattice to form a complete solid solution. Using the FullProf software in pattern matching mode [24] the diffraction pattern was refined in tetragonal single phase structure with *P4mm* (99) space group.

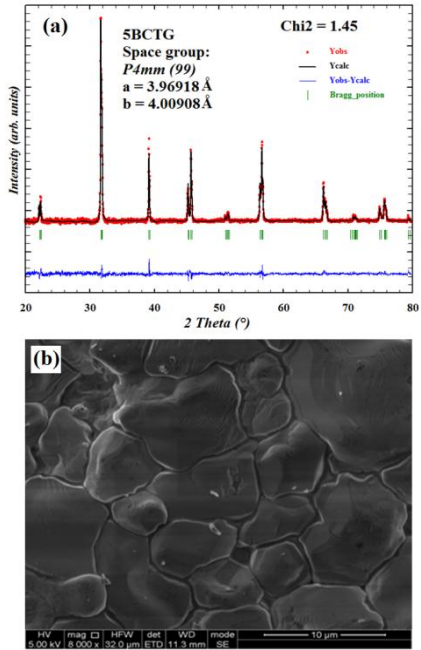

Fig.1: (a) Measured and calculated room temperature X-ray diffraction patterns of BCTG ceramic. The vertical lines show calculated positions of Bragg reflections and the lower curve is the residual diagram performed by using a global profile matching. (b) SEM micrograph of BCTG ceramic.

The SEM micrograph shown in fig. 1(b) was taken at room temperature. A clear grain boundary with grain size of 4~8 μm is observed in BCTG sample. However, the grains are well and strongly connected to each other to form a dense microstructure. Figure 2 displays the dielectric permittivity ε′ measured in a cooling and heating sequences and in a frequency range of 100 Hz to 1 MHz. The BCTG possesses a high dielectric permittivity $\varepsilon_{max}$ of 5130 at the ferroelectric-parraelectric phase transition which occurs at about $T_C$ = 405 K. A small thermal hysteresis of ~4K was observed between the cooling and heating temperature scans. These dielectric data and those of the specific heat $c_p$ measurements shown in the inset of fig. 2 demonstrate that BCTG undergoes a first order phase transition.

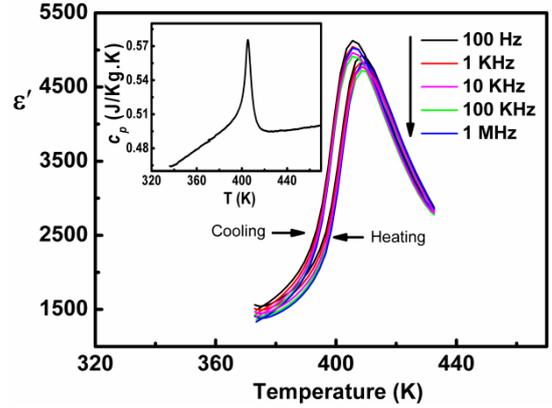

Fig.2: Temperature dependence of permittivity of BCTG ceramic at frequency from $10^2$ to $10^5$ Hz upon cooling after heating scan. The inset presents the temperature dependence of the specific heat capacity.

The hysteresis loops *P–E* measured at six different temperatures are shown in fig. 3(a). The remnant polarization and the coercive field of the sample decrease when the temperature is increased leading to a large-to-slim hysteresis loop transition occurring at $T_C$. The temperature dependence of polarization for different electric fields was deduced from the upper branches of the hysteresis loops and is shown in fig. 3 (b). It is worth to note that for the stronger electric fields and for $T < T_C$, the polarization is increasing with *T* while the Landau theory of the ferroelectrics predicts a monotonically decreasing polarization. A similar result was observed by Moya *et al*. [22]. That could be due to the fact that the sample was not fully polarized under the maximum electric field used in the experiments ($E_{max}$ ~ 4.85 k.V.cm$^{-1}$) leading to upper branches which crossed in the *P(E)* curves.

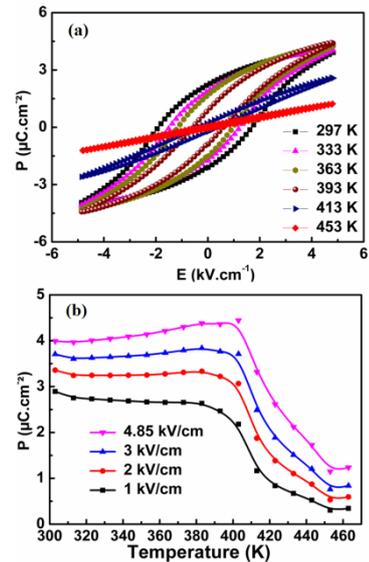

Fig. 3: (a) Electric field dependence of the polarization measured at different temperatures. (b) Temperature dependence of the spontaneous polarization measured at different electric fields.

The temperature dependence of the electrocaloric responsivity of a BCTG ceramic estimated according to eq. (2) is shown in fig. 4 (a). To compare the two methods only the results corresponding to field smaller than 2.4kV.cm$^{-1}$ were presented. The temperature dependence of the heat capacity $c_p(T)$ shown in the inset of fig. 2 has been taken into account. As expected when considering the huge entropy changes concomitant to a phase transition, the temperature corresponding to the maximum of the responsivity was found close to the ferroelectric Curie temperature. Oscillations in $\xi(T)$ are sometime observed in the literature as for instance in the study of Peng *et al.* [25]. Those shown in fig. 4(a) can be explained by one or several of the following causes: i) numerical artifacts arising from the derivatives calculation; ii) the heat capacity $c_p(T)$ was measured in another experiment and just a small mismatch in the temperature scale could result in oscillations; iii) the hysteresis loops were not saturated and as already stated higher, this is causing *P(T)* increasing with temperature for the highest fields and consequently the *P-E* branches were crossing. These errors are giving an overall uncertainty estimated to about $0.05 \cdot 10^{-6}$ K.m.V$^{-1}$.

The pyroelectric coefficient is the change of polarization induced by a change in temperature. This can be measured through the current $i$ flowing in the external circuit and one has [1, 26]:

$$i_E(T) = \left(\frac{\partial P}{\partial T}\right)_E r A \qquad (3)$$

Where *A* is the area of the electrodes and *r* is the heating rate. Using the Maxwell relationship $\left(\frac{\partial T}{\partial E}\right)_T = -\left(\frac{T}{\rho . c_p}\right)\left(\frac{\partial P}{\partial T}\right)_E$, the electrocaloric responsivity can be computed from the pyroelectric current:

$$\xi(T) = -\left(\frac{T}{\rho c_p r A}\right) i_E(T) \qquad (4)$$

In the experiments, in order to reproduce the polarization state that the sample had in the standard method, it would have been necessary to pole the sample to 4.8 kV.cm$^{-1}$. Unfortunately, the used device could only reach 2.4 kV.cm$^{-1}$. Regardless, the following procedure was used. In the step (i) the sample was heated at high temperature $T_H = 453$ K and was poled at the maximum voltage that the generator could yield (2.4 kV.cm$^{-1}$); in the step (ii) thereafter the sample was cooled under this field to $T_L = 333$K; in the step (iii) the field was reduced to a working value $E$ and the sample was heated (5 K.min$^{-1}$) to $T_H$ at the same time the pyroelectric current $i_E(T)$ was measured. The cycle is repeated in step (i) with another value for the working field $E$.

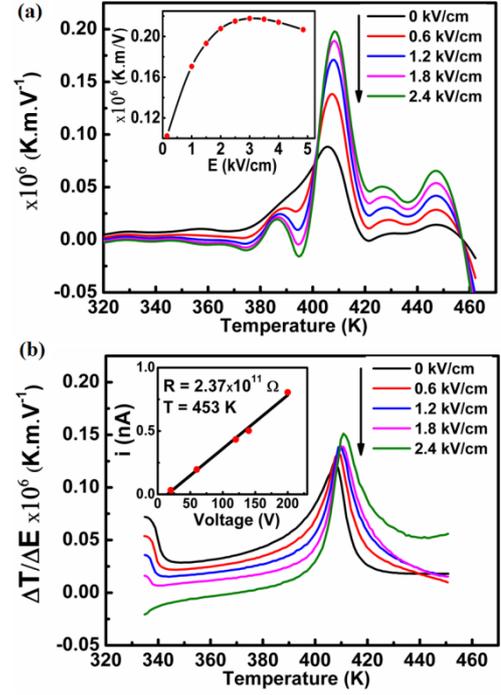

**Fig. 4**: (a) Electrocaloric responsivity estimated from the P(T) dependences at different electric fields according to the Maxwell relation and using the classical method; the inset presents the electrocaloric responsivity maxima dependences on electric fields. (b) Electrocaloric responsivity versus temperature and applied electric field from pyroelectric measurement; the inset presents the as-measured current in function of the working voltage at a fixed temperature (453 K).

The field and temperature dependent electrocaloric responsivity computed from the pyroelectric method is shown in fig. 4(b). At the low temperatures, the responsivity increased slightly until a well defined peak at about the phase transition temperature. Above $T_C$, the sample shown d.c. conduction which contribution was subtracted in a preceding process (inset of fig. 4 (b). It is worth to notice that the $\xi(T)$ curves are ordered from low to high electric field above $T_C$ and conversely below. This behavior can be explained when considering that in the ferroelectric materials, the polarization can be separated into a spontaneous or irreversible and a reversible components [27]:

$$P = P_S + P_{rev} \qquad (5)$$

where $P_{rev} = \varepsilon_0 \chi E$ arises from the polarizability of the material. The dielectric susceptibility $\chi(T, E) = \chi_1 + \chi_2 E + \chi_3 E^2 + \cdots$ being nothing but the slope of the *P-E* curve, examination of the fig. 3 shows that it seems reasonable to assume that the slopes remain almost constant or do not change dramatically along the upper branches of the curves with the noticeable exception for the *P(E)* measured close to $T_C$. So for $T \neq T_C$ one has: $\chi(T, E) \cong \chi_1(T)$. Since $\chi_1 \gg 1$ we take $\chi_1 \approx \varepsilon$, so the measured pyroelectric current defined in eq. (3) is

$$i_E(T) = \frac{\partial P_S}{\partial T} rA + \varepsilon_0 E \frac{\partial \varepsilon}{\partial T} rA \qquad (6)$$

Below $T_C$ one has $(\frac{\partial \varepsilon}{\partial T}) > 0$ when above $T_C$, $(\frac{\partial \varepsilon}{\partial T}) < 0$. Consequently the contribution to the measured current $\varepsilon_0 ErA(\frac{\partial \varepsilon}{\partial T})$ changes sign when the temperature crosses $T_C$ explaining the reverse order of the $i_E(T)$ curves below and above $T_C$. In addition the measured current curves recorded at different $E$ were shifted by values close to the expected ones $\Delta i = \varepsilon_0 \Delta ErA(\frac{\partial \varepsilon}{\partial T})$ where $\Delta E$ is the difference between the working electric fields.

As in the standard method, the observed maxima is close to the phase transition temperature. The maximum value of $\xi$ for the higher field was found 0.15 $10^{-6}$ K.m.V$^{-1}$ which is slightly less than the one found by the *P-E* measurement (0.20 $10^{-6}$ K.m.V$^{-1}$), but in a good agreement taking into account the estimated errors.

In summary, we have elaborated lead free BCT doped germanium Ba$_{0.8}$Ca$_{0.2}$Ti$_{0.95}$Ge$_{0.05}$O$_3$ ceramics to evaluate their pyro and electrocaloric properties. BCTG crystallizes in tetragonal structure with *P4mm* space groups and undergoes a first order phase transition. The electrocaloric responsivity was determined from two different methods based on hysteresis loop measurements and on a new pyroelectric method. Both methods give comparable values for the responsivity within the error of estimation which was found around a mean value of 0.175 $10^{-6}$ K.m.V$^{-1}$. The results indicate the possibility to determine ECE coefficients from a novel pyroelectric approach which is measuring the pyroelectric coefficient directly instead to use sometimes problematic calculation procedure.

∗∗∗

Financial support of this work was provided by the PHC Maghreb grant No. 27958YF. MEM and DM acknowledge CNRST and IOM organizations for the financial support.